# Influence of nanoparticle-graphene separation on the localized surface plasmon resonances of metal nanoparticles

Reza Masoudian Saadabad[1]*, Ahmad Shafiei Aporvari[1], Amir Hushang Shirdel-Havar[2], and Majid Shirdel Havar[3]

**Abstract:** We develop a theory to model the interaction of graphene substrate with localized plasmon resonances in metallic nanoparticles. The influence of a graphene substrate on the surface plasmon resonances is described using an effective background permittivity that is derived from a pseudoparticle concept using the electrostatic method. For this purpose, the interaction of metal nanoparticle with graphene sheet is studied to obtain the optical spectrum of gold nanoparticles deposited on a graphene substrate. Then, we introduce a factor based on dipole approximation to predict the influence of the separation of nanoparticles and graphene on the spectral position of the localized plasmon resonance of the nanoparticles. We applied the theory for a 4 nm radius gold nanosphere placed near 1.5 nm graphene layer. It is shown that a blue shift is emerged in the position of plasmon resonance when the nanoparticle moves away from graphene.

## 1. Introduction

Localized surface plasmon resonances (LSPRs) of metal are due to the coherent oscillation of free electrons and are associated with local enhancement of oscillating electric field supported by metallic nanoparticle. They offer great promise for a range of applications, such as sensor technology (Mock et al. 2003; Schultz 2003; Clark et al. 2007; Lal et al. 2007; Jain and El-Sayed 2008), surface-enhanced Raman scattering (Dahmen and Plessen 2014), and plasmon solar cells (Hagglund et al. 2008). Recently, the combination of metallic nanoparticles and graphene has attracted intense interest due to its great potential in future plasmonic devices (He and Lu 2014). Graphene is a monolayer of carbon atoms and has unique mechanical, electric, and optical properties with a large number of industrial potential applications (Grigorenko et al. 2012; Jastrzębska et al. 2012). Graphene-metal nanocomposites showed great promise for sensors and biosensors (Wu et al. 2010; Kravets et al. 2012) and leaded to noticeable enhancements in the photocurrent and the efficiencies of photovoltaic devices (Chuang et al. 2014).

It is not surprising that the properties of a metallic nanoparticle near a graphene layer different from an isolated nanoparticle due to changes in the electric permittivity of surrounding environment. Since the LSPR wavelengths of metal nanoparticles are highly sensitive to the electric permittivity of the surrounding medium (Vernon et al. 2010), the presence of graphene layer can influence these resonances. The interaction between the LSPR and the graphene sheets observed in some experiments and offers a diverse range of applications (Mueller et al. 2010; Echtermeyer et al. 2011; Hong et al. 2012). It has recently been shown that applying graphene as a substrate for plasmonic metallic nanoparticles offers excellent platforms for biosensors and electronic devices (Polyushkin et al. 2013). Graphene can be used as an ideal spacer for plasmonic metallic nanostructures so that creates a repeatable and stable subnanometer gap for massive plasmonic field enhancements (Mertens et al. 2013; Mcleod et al. 2014).

[1] Department of Physics, University of Sistan and Baluchestan, Zahedan 98135- 674, Iran
[2] Department of Physics, Golestan University, Gorgan, 15759-49138, Iran
[3] Department of Physics, University of Kashan, Kashan, Iran
* E-mail: masoudian_reza@yahoo.com

If metallic nanoparticles are placed near a graphene layer, the resonance wavelength of the nanoparticles will be altered due to particle-graphene interaction (Polyushkin et al. 2013). Knowing the plasmonic response of metallic nanoparticles in the presence of graphene is important for designing nanoparticle-graphene plasmonic systems with a desired resonance frequency. There are some experimental investigations into the effect of graphene layer on the LSPR of metallic nanoparticles (Mertens et al. 2013; Polyushkin et al. 2013; Lee et al. 2014), but there is no general theory to give a physical insight into metal particle-graphene interactions and to predict the dependence of the LSPRs on such interactions.

In this work, we develop a theory of the interaction of graphene sheets with the LSPRs in metal nanoparticles. For this purpose, we consider graphene sheets to be a thin layer with a certain permittivity. Based on the electrostatic approximation the metal nanoparticles on the graphene substrate in a medium are replaced by a homogeneous medium with an effective permittivity in which only the nanoparticles are embedded. This strong method can be used for particles of arbitrary shape. The method makes it possible to avoid intensive numerical calculations and determines the strength of the particle-graphene interaction. In this study, we consider the graphene sheets as a substrate with a specific permittivity. Based on this assumption we calculate the strength of interaction between the nanoparticles and graphene to obtain the optical properties of metallic particle-graphene nanocomposites and predict the spectral position of LSPRs. We then show it is possible to calculate the shifted LSPR wavelength that arises from the distance between the nanoparticles and the graphene using the strength of the nanoparticle-graphene interaction. To this end, a factor is introduced to calculate the reduction of the interaction strength as the distance is increased and to predict the shift of the LSPR wavelength.

**2. Electrostatic resonances of metallic nanoparticles**

The electrostatic approximation can be used for calculating the localized surface plasmon resonances associated with metallic nanoparticles that are much smaller than the wavelength of incident light. The resonances can be studied by the electrostatic eigenmode method introduced by Davis et al (Davis et al. 2009). In this method, the resonance frequencies of a nanoparticle can be calculated using the surface charge distribution $\sigma(\mathbf{r})$ across the surface of the particle that is related to surface charge at a given point on the particle surface $\mathbf{r_q}$ as follows:

$$\sigma(\mathbf{r}) = \frac{\gamma}{2\pi} \oint \sigma(\mathbf{r}_q) \frac{(\mathbf{r}-\mathbf{r}_q)\cdot \mathbf{n}}{|\mathbf{r}-\mathbf{r}_q|^3} dS_q, \qquad (1)$$

where γ is the eigenvalue of the integral equation and takes values $\gamma^k$ each one corresponds to a resonant mode and to a surface charge eigenfunction, $\sigma^k(\mathbf{r})$, of the nanoparticle. There is also similar integral equation for the distribution of surface dipole

$$\tau(\mathbf{r}) = \frac{\gamma}{2\pi} \oint \tau(\mathbf{r}_q) \frac{(\mathbf{r}_q-\mathbf{r})\cdot \hat{\mathbf{n}}_q}{|\mathbf{r}-\mathbf{r}_q|^3} dS_q \quad, \qquad (2)$$

In order to find the eigenvalues and eigenfunctions, the surface of the nanoparticle is divided into a set of small elements so that the surface integral equations (1) and (2) are turned into matrix eigenvalue problem (Mayergoyz et al. 2005; Davis et al. 2009). When the nanoparticle is exposed to an incident radiation with frequency $\omega$, the surface charge distribution of the nanoparticle can be written as the summation of the surface charge eigenmodes

$$\sigma(\mathbf{r},\omega) = \sum_k a^k(\omega)\sigma^k(\mathbf{r}) \tag{3}$$

The expansion coefficient $a^k(\omega)$ is the excitation amplitude of kth resonance mode of the nanoparticle and given by (Mayergoyz et al. 2007)

$$a^k(\omega) = \frac{2\gamma^k \varepsilon_b (\varepsilon(\omega) - \varepsilon_b)}{\varepsilon_b(\gamma^k + 1) + \varepsilon(\omega)(\gamma^k - 1)} \mathbf{p}^k \cdot \mathbf{E} \tag{4}$$

where $\varepsilon(\omega)$ and $\varepsilon_b$ are the dielectric permittivities of metallic nanoparticle and surrounding medium, respectively. In addition, $\mathbf{E}$ and $\mathbf{p}^k$ represent the electric field amplitude of the incident light and the average dipole moment of the particle at resonant frequency $\omega^k$, respectively. It is now obvious that the particle resonates at frequency $\omega^k$ if the real part of denominator of Eq. (4) at this frequency becomes zero. As a result, the resonant frequency of a metal nanoparticle is found using the following equation

$$\varepsilon(\omega^k) = -\varepsilon_b \left( \frac{\gamma^k + 1}{\gamma^k - 1} \right) \tag{5}$$

For single spherical nanoparticle in dipolar approximation $\gamma^k = 3$, hence the resonance frequency is determined by $\varepsilon(\omega^k) = -2\varepsilon_b$.

**3. Particle-graphene interaction**

*3.1 Dielectric function of graphene*

In this paper, we consider graphene to be a layer with certain dielectric function. Defining the dielectric function for graphene sheet enables one to study the optical properties of this two dimensional structure. It has been indicated that the imaginary part of graphene dielectric function shows the optical properties of metal when the electric field of incident light is parallel to graphene sheet, while it shows semiconductor properties for the vertical electric field (Rani et al. 2014). Hence, it is very important to appropriately model graphene to study the optoelectronic properties of graphene-based materials. In this paper, we replace graphene sheets in metallic particle-graphene nanocomposites by a substrate with a specific permittivity. This assumption allows us to apply the electrostatic eigenmode method to predict the shifted plasmon resonances in metallic nanoparticles placed near graphene. Given the nature of the eigenmode method our approach can be extended for any type of nanoparticle of arbitrary shape, although here we focus on spherical metallic nanoparticles.

Graphene is considered to be a uniaxial anisotropic material due to its two-dimensional nature, thus the dielectric function tensor of graphene is written as below

$$\varepsilon_g(\omega) = \begin{bmatrix} \varepsilon_\parallel & 0 & 0 \\ 0 & \varepsilon_\parallel & 0 \\ 0 & 0 & \varepsilon_\perp \end{bmatrix} \tag{6}$$

where $\varepsilon_\perp$ and $\varepsilon_\parallel$ are the normal and parallel components of dielectric function, respectively. The components have been calculated using ab-intio simulation based on density functional theory ( the method has been explained in details in the work of Rani and colleagues (Rani et al. 2014). The parallel component for intrinsic single graphene layer consists of real and imaginary parts (i.e. $\varepsilon_{\parallel} = \varepsilon_{r\parallel} + i\varepsilon_{i\parallel}$) that are wavelength-dependent quantities as shown in Fig. 1, while the normal component can be considered to be a constant quantity, $\varepsilon_{\perp} = 1.2$, for the wavelength of interest. It should be mentioned that the parallel component of permittivity of few-layer graphene is equivalent to that of single layer graphene but the normal component is approximated as the permittivity of graphite, i.e. 2 (Zhu et al. 2013; Du et al. 2014).

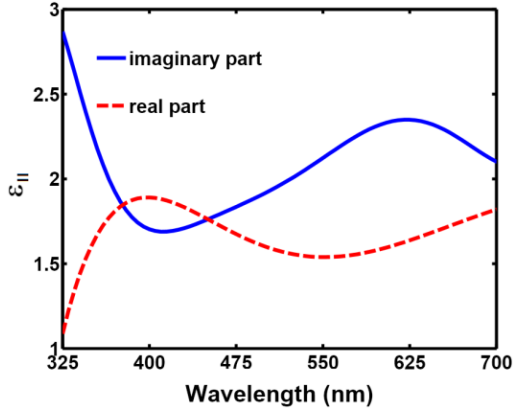

Fig. 1. Imaginary and real parts of parallel component of dielectric function for an intrinsic graphene layer.

*3. 2 Nanoparticle decorated graphene layer*

In this section, we study the plasmon resonance of a metal nanoparticle placed on graphene sheets in a surrounding medium with dielectric permittivity $\varepsilon_b$ (Fig. 2a). We show that the effect of graphene on the LSPR of the nanoparticle can be modeled using pseudoparticle concept. When the particle is placed on graphene, the electric field of the particle's surface charges polarize the graphene layer and thus additional surface charges are induced at interface between the graphene and particle. This phenomenon can be modeled by method of image charges (Jackson 1962; Yamaguchi et al. 1974). In fact, the surface charge of the particle $\sigma(\mathbf{r})$ induces charges on the graphene surface that can be imagined as surface charge $\sigma(\mathbf{r}_1) = ((\varepsilon_b - \varepsilon_g)/(\varepsilon_b + \varepsilon_g))\sigma(\mathbf{r})$ on a pseudoparticle. The electric field from the surface charge of this pseudoparticle now interacts with the nanoparticle. This assumption converts the problem of nanoparticle on graphene into two interacting nanoparticles (Fig. 2b). The excitation amplitude of kth resonance mode of two interacting nanoparticles has been already calculated (McFarland and Duyne 2003; Davis et al. 2009) and now we can use it for the problem of the nanoparticle on graphene layer as bellow:

$$\tilde{a}^k(\omega) = \left( \frac{2\gamma^k \varepsilon_b (\varepsilon(\omega) - \varepsilon_b)}{\varepsilon_b(\gamma^k + 1) + \varepsilon(\omega_k)(\gamma^k - 1) - (\varepsilon(\omega) - \varepsilon_b)\eta T^k} \right) \times \mathbf{p}^k \cdot \mathbf{E} \quad (7)$$

where $\eta = (\varepsilon_b - \varepsilon_g)/(\varepsilon_b + \varepsilon_g)$ and $T^k$ determines overlap between the surface dipole eigenfunction $\tau^k(\mathbf{r})$ of the nanoparticle and the surface charge eigenfunction $\sigma^k(\mathbf{r}_1)$ of the pseudoparticle and is written as follows:

$$T^k = \frac{\gamma^k}{2\pi} \oint\oint \tau^k(\mathbf{r}) \frac{\hat{n}\cdot(\mathbf{r}-\mathbf{r}_1)}{|\mathbf{r}-\mathbf{r}_1|^3} \sigma^k(\mathbf{r}_1) ds\, ds_1 \quad (8)$$

This parameter shows the strength of the interaction between the nanoparticle and the pseudoparticle. Now the denominator of Eq. (7) determines the resonance condition of the system so that we have

$$\varepsilon(\omega^k) = -\varepsilon_b \left( \frac{1 + \eta T^k/(1+\gamma^k)}{1 + \eta T^k/(1-\gamma^k)} \right) \left( \frac{\gamma^k + 1}{\gamma^k - 1} \right) \quad (9)$$

It is easily to understand that Eq. (9) can be written in the same form as Eq. (5) if the surrounding medium with permittivity $\varepsilon_b$ is replaced by a new medium with effective permittivity $\varepsilon_{eff}$ (Vernon et al. 2010)

$$\varepsilon_{eff} = \varepsilon_b \left( \frac{1 + \eta T^k/(1+\gamma^k)}{1 + \eta T^k/(1-\gamma^k)} \right) \quad (10)$$

In another word, defining the effective permittivity helps one to simplify the problem of two interacting nanoparticles to single nanoparticle immersed in a homogeneous medium (Fig. 2c). Overall, as schematically shown in Fig. 2, we replaced the problem of nanoparticle laying on graphene layer in a surrounding medium with the nanoparticle embedded in a homogeneous medium of permittivity $\varepsilon_{eff}$.

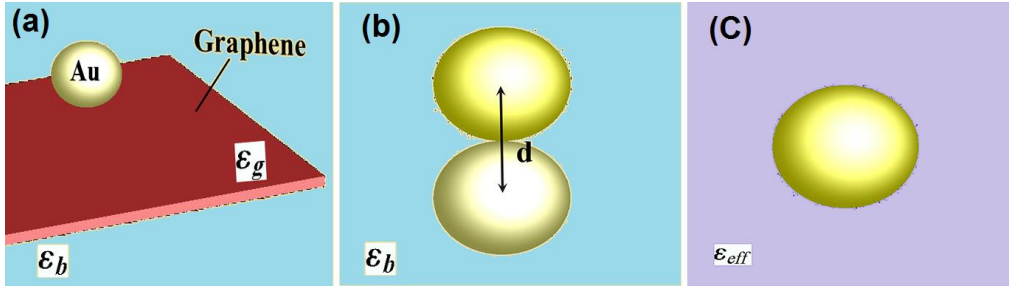

Fig. 2. (a) Particle on graphene layer, where $\varepsilon_g$ and $\varepsilon_b$ are the dielectric permittivity of graphene and surrounding medium. (b) Induced charge on the graphene can be imagined as surface charge on a pseudoparticle that interacts with the particle. (c) The model: nanoparticle in a homogeneous medium with permittivity $\varepsilon_{eff}$.

The theory has been applied to calculate absorption spectrum of a spherical gold nanoparticle (with radius $a = 4$ nm) lying on a few-layer graphene in air ($\varepsilon_b = 1$). Using the electric permittivity of gold has been taken from literature, the resonance frequency of the nanoparticle was determined via Eq. (5) and then the permittivity of graphene at this frequency was used to calculate effective permittivity. The obtained effective permittivity was applied to calculate the cross-section of absorption given by $Q_{abs} = 4ka \, \text{Im}(\varepsilon - \varepsilon_{eff}/\varepsilon + 2\varepsilon_{eff})$ (Fan et al. 2014) and as shown in Fig. 3 there is a strong peak at 524 nm. It is important to note that we did not use the bulk dielectric function of gold in our calculation, but the absorption has been calculated via modified dielectric function (Wormeester et al. 2004) that takes the small size of nanoparticle into account. To test the theory, we have performed finite deference time domain (FDTD) simulations to obtain absorbance using bulk (Johnsons and Christy 1972) and the modified dielectric functions. The optical properties of graphene used in simulations have been taken from Weber et al (Weber et al. 2010). As can be seen from Fig. 3 both simulations show a peak at 528 nm that is in good agreement with what the effective permittivity model predicted. It can be also seen that the modified dielectric function-based calculations predict a broader peak in comparison with one obtained via bulk dielectric function. In fact, broadening of the peak has been observed for gold nanoparticles with a diameter bellow 30 nm (Dinsmore et al. 2001; Wormeester et al. 2004). The prediction of the effective permittivity model and simulations are in agreement with the observed peak position (528 nm) in experiment (Lee et al. 2014).

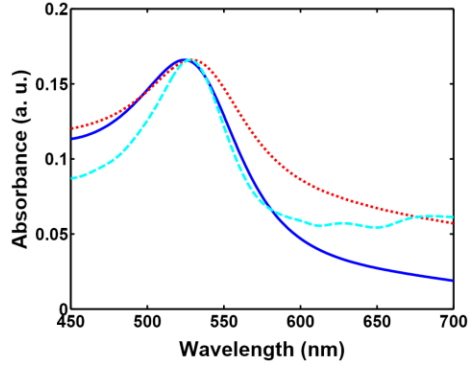

Fig. 3. Absorption spectrum of a gold nanoparticle on few-layer graphene obtained from the effective permittivity model (solid line) and simulation using bulk (dashed line) and modified (dot) dielectric functions of gold.

*3. 3 Distance factor*

Using a series expansion of $|\mathbf{r}-\mathbf{r}_1|^{-3}$ (Vernon et al. 2010; Mortazavi et al. 2012), Eq. (8) can be approximated as follows

$$T^k = \frac{\gamma^k}{2\pi d^3}\left(3(\mathbf{p}^k \cdot \hat{\mathbf{d}})^2 - \mathbf{p}^k \cdot \mathbf{p}^k\right) \tag{11}$$

where $d$ is the distance between the center of the nanoparticle and the pseudoparticle (see Fig. 2b). Here we demonstrate that this approximation can be applied to predict the shift of the resonance wavelength of metallic nanoparticles when the nanoparticles move away from the surface of graphene substrate.

Considering the nanoparticle floating above a graphene layer such that $L$ is the distance from the edge of the nanoparticle to the graphene surface (Fig. 4a). Note that the distance between the center of nanoparticle and pseudoparticle is now $2L+d$ so that Eq. (11) is modified as bellow

$$\Gamma^k = \frac{d^3}{8L^3 + 12L^2d + 6Ld^2 + d^3} \frac{\gamma^k}{2\pi d^3} \times \left[3(\mathbf{p}^k \cdot \hat{\mathbf{d}})^2 - \mathbf{p}^k \cdot \mathbf{p}^k\right] \tag{12}$$

We introduce distance factor $D$ given by

$$D = \frac{d^3}{8L^3 + 12L^2d + 6Ld^2 + d^3} \tag{13}$$

Now Eq. (11) is rewritten as follows:

$$\Gamma^k = D \times T^k \tag{14}$$

The effective permittivity of Eq. (10) can now be used for the system if $T$ factor is substituted with $\Gamma^k$. Distance factor gives a view of how the coupling between the nanoparticle and graphene substrate can be decreased with the increase of graphene-nanoparticle separation. In fact, as the nanoparticle moving away from graphene surface the value of $\Gamma^k$ is decreased and thus a reduction in the effective permittivity of the system is occurred that, in turn, leads to a blue shift in the resonance position of the particle.

The dependence of the optical response of a gold nanoparticle supported by graphene sheets on the separation has been investigated using distance factor and shown in Fig. 4. The figure includes a schematic of the system structure. Fig. 4(b) shows the quantity of $1-Q_{extinction}$, which is proportional to the transmission value (Niu et al. 2012), as a function of wavelength. Extinction cross-section $Q_{extinction}$ has been calculated via the summation of the cross-section of scattering and absorption ($Q_{abs}+Q_{sca}$) (Fan et al. 2014). As expected for particles of less than 40 nm of diameter, the calculations showed that the extinction is completely dominated by absorption and scattering can be neglected. As can be seen from the figure, there is a blue shift of dip position as the separation of the nanoparticle and graphene becomes wider. This phenomenon was already observed in experiment (Niu et al. 2012). Fig. 4(c) shows a comparison between shifts predicted by effective permittivity model and FDTD simulation. It is clear that a 10.5 nm blue shift is predicted by both methods with increasing the separation from 0 to 5 nm.

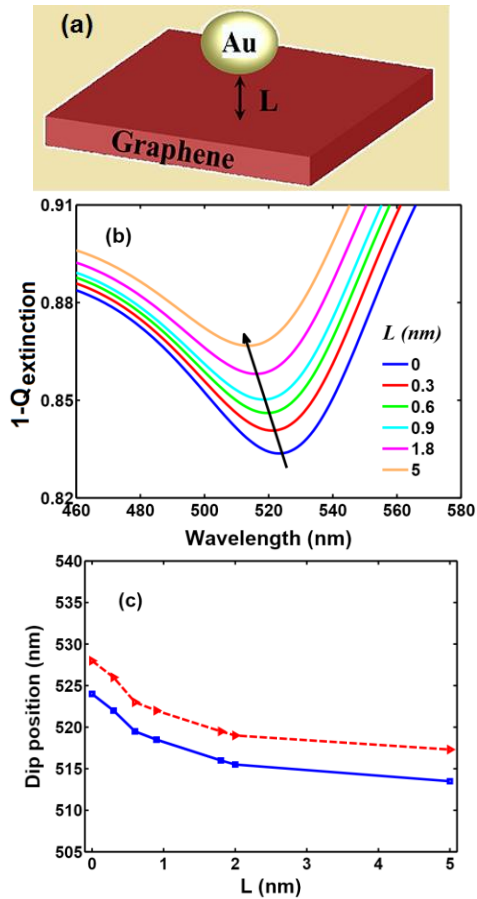

Fig. 4 : (a) Nanoparticle above a graphene layer. (b) Dependence of resonance on gold nanoparticle-graphene separation. The arrow shows the movement of dip position as the gap between nanoparticle and graphene becomes wider. (c) Calculated shifts of dip position using the effective permittivity (solid line) and FDTD simulation (dashed line).

The blue shift can be explained by the value of factor $\Gamma^k$. Increasing separation of graphene and nanoparticle decreases $|\Gamma^k|$ (and thus effective permittivity) and consequently leads to a blue shift in dip position. The wavelength position of the dip and the value of $\Gamma^k$ versus the separation are shown in Fig. 5. It can be seen that for large separation the value of $\Gamma^k$ is almost zero and thus the value of effective permittivity can be approximated as $\varepsilon_b$. As a result, increasing the separation beyond 15 nm causes no further shift in dip position.

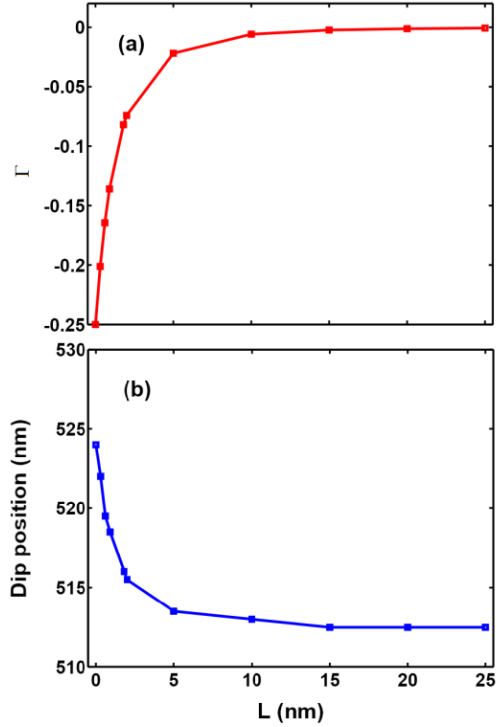

Fig. 5. (a) Dependence of $\Gamma^k$ and dip position on gold nanoparticle-graphene separation.

## 4. Conclusion

A theory based on electrostatic approximation has been developed to predict the shift of the LSPRs of metallic nanoparticles placed near graphene sheets. The graphene alters the electric permittivity of surrounding medium of the particles that, in turn, leads to a change in the LSPR wavelength. The changed LSPR has been calculated via introducing an effective permittivity that is dependent on the interaction between graphene layer and metal nanoparticles dispersed on. We have then derived a parameter that determines the strength of the interaction when there is a gap between graphene and the particles and gives a view of how the resonance wavelength is changed. We have shown that the results of our method are in good agreement with numerical simulations. We believe that the theory can be useful for designing nanoparticle-graphene plasmonic systems.

**References**

Chuang MK, Lin SW, Chen FC, Chu CW, Hsu CS (2014) Gold nanoparticle-decorated graphene oxides for plasmonic-enhanced polymer photovoltaic devices. Nanoscale 6:1573–1579.

Clark AW, Sheridan AK, Glidle A, Cumming DRS, Cooper JM (2007) Tuneable visible resonances in crescent shaped nano-split-ring resonators. Appl Phys Lett 91:3109.

Dahmen C, Plessen G von (2014) Control of Gold Nanostructure Morphology by Variation of Temperature and Reagent Ratios in the Turkevich Reaction. Aust J Chem 68:858–862.

Davis TJ, Vernon KC, Gómez DE (2009) Designing plasmonic systems using optical coupling between nanoparticles. Physical Review B 79:155423.

Dinsmore AD, Prasad ER, Prasad V, Levitt AC, Weitz DA (2001) Three-dimensional confocal microscopy of colloids. Appl Opt 40:4152–4159.

Du W, Hao R, Li E-P (2014) The study of few-layer graphene based Mach−Zehnder modulator. Optics Communications. OPt Comm 323:49–53.

Echtermeyer TJ, Britnell L, Jasnos PK, Lombardo A, Gorbachev R V., Grigorenko AN, Geim AK, Ferrari AC, Novoselov KS (2011) Strong plasmonic enhancement of photovoltage in graphene. Nat Commun 22:458.

Fan X, Zheng W, Singh DJ (2014) Light scattering and surface plasmons on small spherical particles. Light: Science & Applications 3:179.

Grigorenko AN, Polini M, Novoselov KS (2012) Graphene plasmonics. Nat Photon 6:749–758.

Hagglund C, Zach M, Petersson G, Kasemo B (2008) Electromagnetic coupling of light into a silicon solar cell by nanodisk plasmons. Appl Phys Lett 92:053110.

He X, Lu H (2014) Graphene-supported tunable extraordinary transmission. Nanotechnology 25:325201.

Hong H, Yang K, Zhang Y, Engle JW, Feng LZ, A.Yang Y, Nayak TR, Goel S, Beam J, Theuer CP, Barnhart TE, Liu Z, Cai WB (2012) In Vivo Targeting and Imaging of Tumor Vasculature With Radiolabeled, Antibody-Conjugated Nanographene. ACS Nano 6:2361–2370.

Jackson JD (1962) Classical Electrodynamics. John Wiley and Sons, New York

Jain PK, El-Sayed MA (2008) Noble Metal Nanoparticle Pairs: Effect of Medium for Enhanced Nanosensing. Nano Lett 8:4347–4352.

Jastrzębska AM, Kurtycz P, Olszyna AR (2012) Recent advances in graphene family materials toxicity investigations. Journal of nanoparticle research 14:1320.

Johnsons PB, Christy RW (1972) Optical constants of nobel metals. Phys Rev B 6:4370–4379.

Kravets VG, Schedin F, Jalil R, Britnell L, Novoselov KS, Grigorenko AN (2012) Surface Hydrogenation and Optics of a Graphene Sheet Transferred onto a Plasmonic Nanoarray. J Phys Chem C 116:3882–3887.

Lal S, Link S, Halas NJ (2007) Nano-optics from sensing to waveguiding. Nat Photon 1:641–648.

Lee J, Kim J, Ahmed SR, Zhou H, Kim J-M, Lee J (2014) Plasmon-Induced Photoluminescence Immunoassay for Tuberculosis Monitoring Using Gold-Nanoparticle-Decorated Graphene. ACS Applied Materials & Interfaces 6:21380–21388.

Mayergoyz ID, Fredkin DR, Zhang Z (2005) Electrostatic (plasmon) resonances in nanoparticles. Phys Rev B 72:155412.

Mayergoyz ID, Zhang Z, Miano G (2007) Analysis of Dynamics of Excitation and Dephasing of Plasmon Resonance Modes in Nanoparticles. Phys Rev Lett 98:147401.

McFarland AD, Duyne R Van (2003) Single Silver Nanoparticles as Real-Time Optical Sensors with Zeptomole Sensitivity. Nano Lett 3:1057–1062.

Mcleod A, Vernon KC, Rider AE, Ostrikov K (2014) Optical coupling of Au nanoparticles on vertical graphenes to maximize SERS response. Optics Letters 39:2334–2337.

Mertens J, Eiden AL, Sigle DO, Huang F, Lombardo A, Sun Z, Sundaram RS, Colli A, Tserkezis C, Aizpurua J, Milana S, Ferrari AC, Baumberg JJ (2013) Controlling Subnanometer Gaps in Plasmonic Dimers Using Graphene. Nano Lett 13:5033–5038.

Mock JJ, Smith DR, Schultz S (2003) Local Refractive Index Dependence of Plasmon Resonance Spectra from Individual Nanoparticles. Nano Lett 3:485–491.

Mortazavi D, Kouzani AZ, Vernon KC (2012) A RESONANCE TUNABLE AND DURABLE LSPR NANO-PARTICLE SENSOR: AL2O3 CAPPED SILVER NANO-DISKS. Progress In Electromagnetics Research 130:429–446.

Mueller T, Xia FNA, Avouris P (2010) Graphene photodetectors for high-speed optical communications. Nat Photon 4:297–301.

Niu J, Shin YJ, Lee Y, Ahn J, Yang H (2012) Graphene induced tunability of the surface plasmon resonance. Appl Phys Lett 100:061116.

Polyushkin DK, Milton J, Santandrea S, Russo S, Craciun MF, Green SJ, Mahe L, Winolve CP, Barnes WL (2013) Graphene as a substrate for plasmonic nanoparticles. Journal of Optics 15:114001.

Rani P, Dubey GS, Jindal VK (2014) DFT study of optical properties of pure and doped graphene. Physica E: Low-dimensional Systems and Nanostructures 62:28–35.

Schultz DA (2003) Plasmon resonant particles for biological detection. Curr Opin Biotechnol 14:13–22.

Vernon KC, Funston AM, Novo C, Gómez DE, Mulvaney P, Davis TJ (2010) Influence of particle-substrate interaction on localized plasmon resonances. Nano lett 10:2080–6.

Weber JW, Calado VE, Sanden MCM Van De (2010) Optical constants of graphene measured by spectroscopic ellipsometry. Applied Physics Letters 97:091904.

Wormeester H, Kooij ES, Poelsema B (2004) Self-Assembled Thin Films: Optical Characterization. In: Dekker Encyclopedia of Nanoscience and Nanotechnology. pp 3361–3371

Wu L, Chu HS, Koh WS, Li EP (2010) Highly sensitive graphene biosensors based on surface plasmon resonance. Opt Express 18:14395.

Yamaguchi T, Yoshia S, Kinbara A (1974) Effect of retarded dipole-dipole interactions between island particles on the optical plasma-resonance absorption of a silver-island film. Thin Solid Films 21:173–187.

Zhu B, Ren G, Zheng S, Lin Z, Jian S (2013) Nanoscale dielectric-graphene-dielectric tunable infrared waveguide with ultrahigh refractive indices. Opt Express 21:17089–17096.